# Attempt to Generate Narrow Linewidth, CW Terahertz Radiation by Using Optical Frequency Comb

S. M. Iftiquar, Kiyomi Sakai, Masahiko Tani,
Bambang Widiyatmoko, Motonobu Kourogi, Motoichi Otsu

*Abstract-* We report on an attempt to generate highly stable continuous terahertz (THz) wave by using optical frequency comb (OFC). About 10-nm wide OFC has been generated through a deep phase modulation of a 852 nm laser line in lithium niobate crystal cavity. The multiple optical modes (side bands) of the OFC, which are equally separated from each other by the modulation frequency (=6 GHz) are taken as the frequency reference. When another semiconductor laser is frequency locked, the stability of the difference frequency between the master laser and the second laser is improved on the same order of the RF modulator. An ultra-narrow line and tunable THz radiation source can be achieved by photomixing of this stable difference-frequency optical beat in a photoconductive antenna.

## I. INTRODUCTION

Generation of submillimeter wavelength (THz) radiation through photomixing, or optical-heterodyne conversion of two single mode laser beams in a photoconductive antenna [1] is a promising approach in realizing a stable and tunable THz radiation source. The stability (linewidth) and tunability of this kind of radiation source is mostly determined by those of the optical pump source. Semiconductor lasers are good candidates as the excitation laser source because of its wide tunability, compactness, and the ease for the operation. In addition, the advanced diode laser technology is already available at hand. For example, two external grating-stabilized diode lasers were used as the photomixing source, and a wide tunability (0.6-3.2 THz) and a linewidth of ~5 MHz was achieved [2]. A Fabry-Perot cavity was used to further stabilize the diode lasers, and a linewidth about 1 MHz was achieved [3]. The Fabry-Perot cavity shows a optical frequency comb in transmission spectrum, whose mode separation corresponds to the FSR (Free Spectral Range) of the cavity. However, the linewidth of a mode ($\Delta\nu$), which is determined by the finesse of the Fabry-Perot cavity ($\Delta\nu$=FSR/finesse), is usually on a MHz order. The modes in such a passive optical device are also subject to mechanical and temperature instabilities.

With the recent advancement in the opto-electronics we can now generate an optical frequency comb by a deep phase modulation of a continuous-wave (cw) laser beam [4]. For example, some of the co-authors demonstrated an optical frequency comb generation with a bandwidth of 4 THz (mode separation=5.8 GHz) and a mode stability about 30 Hz at 1.55 μm line from a diode laser by using a LiNbO$_3$ crystal cavity as the phase modulator [5]. In this work we have tried to stabilize two diode lasers by using an optical frequency comb (OFC).

The OFC is a result of a deep phase modulation of a laser beam and consists of many sidebands around the carrier line due to the phase modulation. It is equivalent to phase modulated (or phase locked) pulses, whose repetition frequency is corresponding to the RF modulation frequency (twice of it). Once such an OFC is achieved, a second laser can be frequency or phase locked to one of the OFC sideband. The difference frequency between the second laser and the master laser used for OFC (or a third laser locked to a different mode of the OFC) is stabilized to the order of the stability obtained by the RF modulator, for which a stability of $\Delta\nu/\nu = 10^{-10}$ is easily achieved. In addition to the good stability, the absolute difference frequency ($\nu_{dif}$) can also be determined by the mode difference number (k) and the off-set frequency of the laser line to a mode of the OFC ($\nu_{offset}$): $\nu_{dif} = \nu_{offset} + k\nu_{mod}$, where $\nu_{mod}$ is the modulation frequency. The mode difference number k can be determined from the relation that $\Delta\nu_{dif} = k \Delta\nu_{mod}$, where $\Delta\nu_{mod}$ is the variation in the modulation frequency and $\Delta\nu_{dif}$ is the change in the difference frequency (both are measurable values).

The narrow line and tunable THz radiation source is desired for the high-resolution molecular spectroscopy, because the vibrational and rotational transition in molecules or molecular radicals are observed in the THz frequency range. In astro-physics, there is a big demand for a compact and stable THz local oscillator for the heterodyne detection of millimeter and submillimeter wavelength radiation, for example, in VLBI [6] or satellite observatories for precise positioning of heavenly bodies and spectroscopy of interstellar molecular gases, respectively.

S. M. Iftiquar, K. Sakai and M. Tani are with the Terahertz Optoelectronic Research Group, Kansai Advanced Research Center, CRL (MPT), 588-2 Iwaoka, Iwaoka-cho, Nishi-ku, Kobe, Hyogo 651-2401, Japan
B. Widiyatmoko, M. Kourogi and M. Otsu are with the Interdisciplinary Graduate School of Science and Technology, Tokyo Institute of Technology, 4259 Nagatsuta, Midori-ku, Yokohama, kanagawa 226, Japan, & Kanagawa Academy of Science and Technology, KSP East building,
Room 408, 3-2-1 Sakado, Takatsu-ku, Kawasaki, Kanagawa 213, Japan

## II. : OPTICAL FREQUENCY COMB GENERATOR (OFCG)

A 852-nm diode laser beam is phase modulated in a lithium niobate (LiNbO$_3$) crystal with a monolithic Fabry-Perot cavity. One of the cavity mirrors is a plane mirror and the other is a spherical mirror with a radius of curvature of 50 mm. Details of the OFCG can be obtained in [7]. In our experiment the crystal length is 22 mm (along z-axis) with two high reflecting mirrors at the two ends whereas gold electrodes are deposited on two side surfaces normal to the c-axis. At 852-nm wavelength the refractive index of O-ray obtained through a Sellmeier equation and experimental data are 2.2410 and 2.2465, respectively. Thus, the obtained FSR is 3.03 GHz. Power reflectivity of the mirrors is 99.6 %. With this crystal cavity, a 10-watt RF power at 6.06 GHz has been applied to phase modulate the laser beam. A number of frequency sidebands are generated, which can briefly be described by the following equation in an ideal condition,

$$E_{pm} = E_0 \left\{ \sum_{k \geq 0} J_k(p) e^{ik\Omega t} + \sum_{k>0} (-1)^k J_k(p) e^{-ik\Omega t} \right\} e^{i\omega t}, \quad (1)$$

where, $J_k$ Bessel function of order k, $\Omega$ RF modulation frequency, p modulation depth.

The generated OFC with a band width of about 10 nm is shown in Fig. 1(a). Presently the used laser power is of 10 mW and obtained OFC output power is about 1 mW.

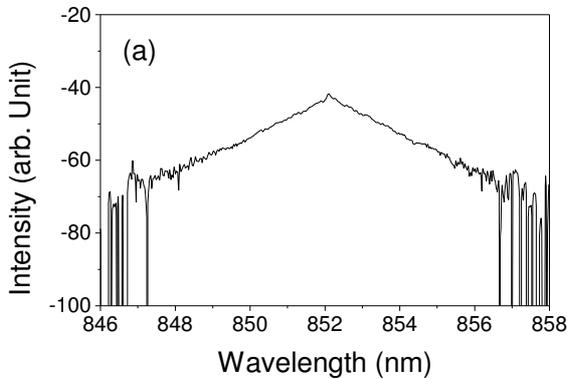

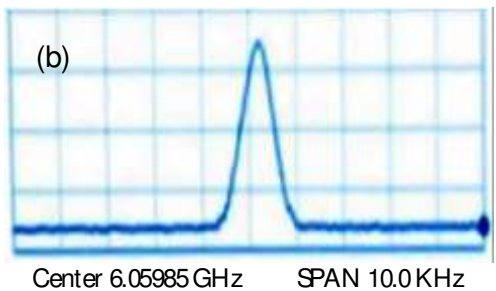

Fig.1(a): OFC generated around 852 nm carrier laser line, (b) Beat signal of OFC sidebands detected through RF spectrum analyzer.

The stability of the RF oscillator is 2x10$^{-7}$. This gives a RF frequency linewidth of 1.2 kHz at 6 GHz, which was confirmed by measuring the RF signal at 6.06 GHz as shown in Fig. 1(b). The stability of an optical beat between two modes in OFC is estimated from this RF oscillator stability. For mode difference $\Delta k=100$, the beat frequency is 606 GHz and the linewidth of the beat node is expected to be 121 kHz.

## III. LASER STABILIZATION

We used two external-cavity diode lasers with a grating mirror on one end (Littrow mount), both of which can be tuned at around 852 nm. The first laser (LD1), that is used for OFC generation, has been frequency stabilized with a reference Fabry-Perot cavity. The finesse of the cavity is about 530 (FSR ~10 GHz). A low voltage modulation is given to a PZT element on one of the cavity mirrors, and transmission of the beam is detected by a photodetector. This signal has been used to frequency stabilize the LD1. This stabilization scheme is shown in Fig. 2. Since the RF oscillator frequency stability ($\Delta\nu/\nu$) is $2\times10^{-7}$, the difference frequency noise at the laser line (852 nm) of the first sideband will be the same as the RF frequency stability and is expected to be about 1 kHz. For the 165-th order sideband that corresponds to about 1 THz frequency separation from the carrier frequency (352 THz), the frequency noise will be about 165 kHz in an ideal case. This noise will be reflected in the emitted THz radiation as well.

A schematic diagram of the system for offset locking is shown in Fig. 2. At the photodiode PD2 a beat frequency is generated, which corresponds to the frequency separation between the LD2 and one of the OFC sideband. Since the estimated power of the 5 th sideband of the OFC is only 2.6 µW, a higher power of the LD2 beam (about 5 mW out of 80 mw total power) is used to generate more intense beat signal. According to equation below the intensity of the beat signal depends on power of both the difference frequency components:

$$P_i = \eta \left\{ P_0 + 2(mP_1P_2)^{1/2} \cos(\omega_1 - \omega_2)t \right\}, \quad (2)$$

where, $P_i$ PD output power, $\eta$ PD efficiency (~45%@850nm), $P_0=P_1+P_2$, $P_1$ OFC sideband power, $P_2$ power of LD2 beam used, $\omega_1$ OFC side band frequency, $\omega_2$ LD$_2$ frequency, $m$ mixing efficiency (~50%). Thus, ~75µW beat signal is expected to be generated at the PD2 photodiode, which is then amplified through a 30-dB preamplifier to detect the signal by a RF spectrum analyzer. This amplified beat signal is tuned to about 95 MHz by the PZT actuator mounted to the feed back grating of the LD2. Using a 100-MHz local oscillator with a –30-dBm output power, an intermediate frequency between the beat signal and the local oscillator is generated in a double balanced mixer (DBM). The signal of the intermediate frequency from the DBM is then passed through an IF-voltage converter and low noise amplifier before feeding it to a lock in amplifier (LIA).

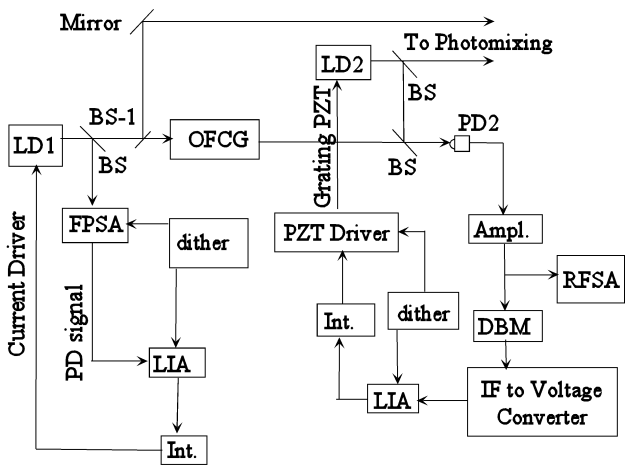

Fig.2: Schematic diagram of laser stabilization. BS; 90:10(T:R) beam splitter, BS-1; 40:60 beam splitter, LIA; lock in amplifier, Int.; Integrator, Ampl.; low noise amplifier, FPSA; Fabry-Perot spectrum analyzer, RFSA; RF spectrum analyzer, DBM; double balanced mixer, OFCG; optical frequency comb generator.

The LIA then generates the corresponding signal, which is fed back to the PZT actuator attached on grating of the LD2. The correction signal is generated when frequency of LD2 is modulated, through current modulation, of about 10 kHz and fed to the LD2.

### IV. THz Generation

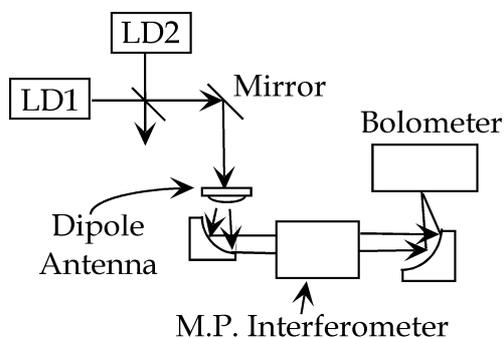

Fig. 3: Schematics of generation of THz radiation through Photomixing and detection through M.P. interferometer and Bolometer combination.

Major power of the laser beam from LD1 and LD2 are spatially coupled for photomixing. A 50:50 beam splitter/combiner has been used for this purpose, as shown in Fig. 3. A long dipole antenna with 1 mm antenna length and 5 μm antenna gap, fabricated on a low-temperature-grown GaAs, is used for the photomixing. A DC voltage bias is applied across the antenna gap. The generated THz radiation is then measured through a Martin-Puplett polarizing interferometer [8] with a liquid-helium cooled (4.2 K) InSb hot-electron bolometer.

Unfortunately, one of the lasers had a break-down during the experiment. The spectrum of the THz radiation is not available yet at this stage of our experiment. The data for the THz radiation spectrum will be presented at the Conference.

### V. Summary

We have demonstrated stabilization of two lasers for its relative frequency difference by using an optical frequency comb. With this laser system an optical beat and thus THz frequency radiation with the same stability of the RF oscillator used is achievable through the photomixing in a photoconductive antenna. Ultimate frequency stability better than 1 kHz is feasible at THz frequencies by using a RF oscillator with a stability of $10^{-10}$ (already commercially available) and the phase locking of the lasers to the OFC. The advantages of this system, in addition to the stability, are the tunability and the relative ease in determining the absolute frequency, which can be calculated from the number of mode difference and the off-set frequency.